\documentclass[conference]{IEEEtran}
\usepackage[T1]{fontenc}
\usepackage{float}
\usepackage{dblfloatfix}
\usepackage[dvips]{graphicx}
\graphicspath{{../eps/}}
\DeclareGraphicsExtensions{.eps}
\ifCLASSINFOpdf
\else
\fi

\ifCLASSOPTIONcompsoc
  \usepackage[caption=false,font=normalsize,labelfont=sf,textfont=sf]{subfig}
\else
  \usepackage[caption=false,font=footnotesize]{subfig}
\fi

\usepackage{amsmath}
\usepackage[cmintegrals]{newtxmath}
\usepackage{graphicx}
\newtheorem{theorem}{Theorem}

\newtheorem{lemma}{Lemma}
\begin{document}

\title{Downlink NOMA in Multi-UAV Networks over Bivariate Rician Shadowed Fading Channels}

\author{
\IEEEauthorblockN{ Tan Zheng Hui Ernest*, A S Madhukumar*, Rajendra Prasad Sirigina*, Anoop Kumar Krishna**}
\IEEEauthorblockA{*School of Computer Science and Engineering\\
Nanyang Technological University, Singapore\\
Email: tanz0119@e.ntu.edu.sg, \{raje0015, asmadhukumar\}@ntu.edu.sg}
\IEEEauthorblockA{**Airbus Singapore Pte Ltd, Singapore\\
Email: anoopkumar.krishna@airbus.com}
}

\maketitle

\begin{abstract}
Unmanned aerial vehicles (UAVs) are set to feature heavily in upcoming fifth generation (5G) networks. Yet, the adoption of multi-UAV networks means that spectrum scarcity in UAV communications is an issue in need of urgent solutions. Towards this end, downlink non-orthogonal multiple access (NOMA) is investigated in this paper for multi-UAV networks to improve spectrum utilization. Using the bivariate Rician shadowed fading model, closed-form expressions for the joint probability density function (PDF), marginal cumulative distribution functions (CDFs), and outage probability expressions are derived. Under a stochastic geometry framework for downlink NOMA at the UAVs, an outage probability analysis of the multi-UAV network is conducted, where it is shown that downlink NOMA attains lower outage probability than orthogonal multiple access (OMA). Furthermore, it is shown that NOMA is less susceptible to shadowing than OMA. 
\end{abstract}


\IEEEpeerreviewmaketitle

\section{Introduction}

In upcoming fifth generation (5G) networks, the application of unmanned aerial vehicles (UAVs) is expected to feature heavily for a multitude of roles, including as aerial base stations \cite{yadav2018full}. However, spectrum is a scarce resource in UAV communications, with UAV control links sharing the spectrum together with other existing systems \cite{matolak2017air_suburban}.

To improve spectrum utilization in 5G UAV communications, power domain non-orthogonal multiple access (NOMA) can be adopted to accommodate more downlink UAVs in multi-UAV networks than orthogonal multiple access (OMA) schemes. A typical scenario for multi-UAV networks with downlink NOMA involves the ground station (GS) using superposition coding to transmit signals simultaneously to the downlink UAVs. Thereafter, UAVs with strong channel gains employ successive interference cancellation (SIC) to recover the desired signal while UAVs with weak channel gains decode the desired signal by treating other UAVs' messages as noise, i.e., interference-ignorant (II) detection. 

Since being considered as a candidate technology for 5G, NOMA has been widely investigated in the literature, particularly for cellular networks in terms of reliability \cite{yang2016general,cui2016novel,kim2015non,zhang2017downlink}. For instance, the outage probability at downlink nodes with NOMA was analyzed for Rayleigh fading channel models in \cite{yang2016general,cui2016novel,kim2015non,zhang2017downlink} with fixed power allocation schemes \cite{kim2015non} and dynamic power control schemes \cite{yang2016general,cui2016novel,zhang2017downlink}. One of the main observations in \cite{yang2016general,cui2016novel,zhang2017downlink} was that NOMA can achieve a lower outage probability than OMA. However, the extent to which these observations can be applied to multi-UAV networks is currently unclear due to a difference in the operating environment. 

One of the main difference between cellular networks and multi-UAV networks is the modeling of node locations. In this regard, stochastic geometry has been widely studied in the literature for the purpose of accurate system analysis. For cellular systems, the Poisson point process (PPP) model has been widely applied \cite{chetlur2017downlink}. However, for multi-UAV networks, the PPP model is unsuitable given a fixed number of UAVs operating in a multi-UAV network \cite{chetlur2017downlink}, e.g., when UAVs are deployed as aerial BSs \cite{chetlur2017downlink,wang2018modeling}. Instead, one can use the homogeneous binomial point process (BPP) model for the spatial locations of UAVs \cite{chetlur2017downlink,wang2018modeling}. In this spirit, it is noted that the application of the BPP model for multi-UAV networks with downlink NOMA has not yet received much attention in the literature.

In addition to the modeling of UAV locations, UAV channel models can also differ from those commonly used for cellular systems. As an example, it is noted that, apart from Rayleigh fading, Rician and Rician shadowed fading \cite{matolak2017air_suburban,tan2018joint,tan2018ricianShad} can also be encountered in UAV communications. In particular, the reliability of interference-limited UAV communications was studied in \cite{tan2018joint}. Using a power series approach, the non-centered Chi-squared probability density function (PDF) was expressed as a power series to enable derivations of closed-form outage probability expressions. Utilizing the same power series approach, the relevant PDF and cumulative distribution function (CDF) expressions for a Rician shadowed fading model was presented in closed-form \cite{tan2018ricianShad}. The closed-form solutions in \cite{tan2018ricianShad} was subsequently applied towards outage probability analysis of UAV communications in a Rician shadowed fading environment. 

Thus far, the studies in \cite{tan2018joint} and \cite{tan2018ricianShad} have only considered univariate fading models, which are suitable towards modeling point-to-point links in OMA-based systems. However, in the context of NOMA, univariate fading models are unsuitable due to dependent UAV links, which can arise due to similar operating environments. To overcome such a limitation, a bivariate Rician shadowed fading model can be considered. In this aspect, a semi-analytical expression for the PDF of the bivariate Rician shadowed fading model was recently presented in \cite{lopez2018bivariate}. However, the semi-analytical nature of the PDF may not lend itself to a tractable computation of common performance metrics, e.g., outage probability. 

Therefore, in this paper, we present closed-form expressions for the PDF and marginal CDFs of the bivariate Rician shadowed fading model. Using these expressions, the reliability of downlink NOMA in a multi-UAV network is analyzed for bivariate Rician shadowed fading channels. In particular, the reliability of downlink NOMA in UAV communications is compared against conventional OMA operating over univariate Rician shadowed fading channels. The major contributions of this paper are as follows.
\begin{itemize}
	\item Closed-form expressions for the joint PDF and marginal CDFs are presented for the bivariate Rician shadowed fading model using a power series approach.
	\item Using the joint PDF and marginal CDF expressions, closed-form outage probability expressions are derived for downlink NOMA in a multi-UAV network. It is shown that the multi-UAV network with downlink NOMA attains lower outage probability than OMA and is less affected by the impact of shadowing. 
\end{itemize}

The organization of this paper is as follows. The system model is introduced in Section \ref{sec_sys_model}. In Section \ref{sec_bi_var}, the bivariate Rician shadowed fading model is discussed, with outage probability expressions presented in Section \ref{sec_outage}. Numerical results are discussed in Section \ref{sec_num_res} before the conclusion of the paper in Section \ref{sec_conclusion}.

\section{System Model} \label{sec_sys_model}
\begin{figure} [tpb]
\centering
\includegraphics [width=0.65\columnwidth]{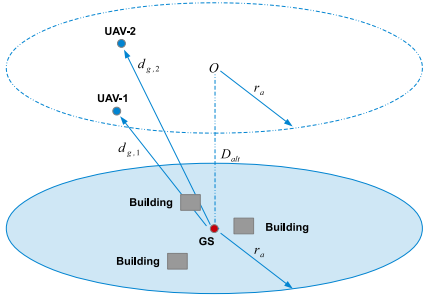} 
\caption{A multi-UAV network with downlink NOMA in a suburban environment is illustrated here. The downlink UAVs, UAV-1 and UAV-2, simultaneously communicate on the same spectrum with the GS.}
\label{fig:block_diagram}
\end{figure}

We consider a multi-UAV network with downlink NOMA that is operating in a suburban environment, as depicted in Fig. \ref{fig:block_diagram}. Without loss of generality, we consider the specific case of a single-antenna GS that is simultaneously transmitting downlink data to two single-antenna downlink UAVs, UAV-1 and UAV-2, on the same spectrum. To model the deployment of the downlink UAVs, the spatial locations of the UAVs are assumed to follow a BPP \cite{chetlur2017downlink, wang2018modeling}. Further, it is assumed that the downlink UAVs are operating at the same altitude ($D_{alt}$) to comply with altitude restrictions. 

As the multi-UAV network operates in a suburban setting, the Rician shadowed fading channel model has been shown to be suitable towards modeling fading and shadowing in UAV communications \cite{tan2018ricianShad}. However, unlike in \cite{tan2018ricianShad}, we assume bivariate Rician shadowed fading channels between the UAVs and the GS to account for dependency of the links due to similar UAV channel conditions. Finally, Doppler shift is assumed to be compensated in this work \cite{tan2018joint}.

\subsection{Distance Distribution of the Downlink UAVs}

Following the work in \cite{chetlur2017downlink}, let the spatial location of the downlink UAVs be uniformly distributed in a disc with radius $r_a$, angle $\left[0,2\pi\right)$, and origin $O$ that is above the GS at an altitude of $D_{alt}$. With the GS directly below the origin $O$, the Euclidean distance (km) between UAV-$i$ and the GS is given as $d_{g,i}=\sqrt{D_{g,i}^2 + D_{alt}^2}$, where $D_{g,i}$ is the Euclidean distance from the GS that is computed using the projection of UAV-$i$ onto the ground plane. As the spatial location of downlink UAV-$i$ follows a BPP, the PDF $f_{d_{g,i}}(w)$ of $d_{g,i}$ is given as $f_{d_{g,i}}(w) = \frac{2w}{r_a^2}$ \cite[eq. (3)]{chetlur2017downlink}, where $\sqrt{D_{alt}^2+\lambda_{g,i}^2} \leq w \leq \sqrt{D_{alt}^2+\lambda_{g,i}^2+r_{a}^2}$, $0 < \lambda_{g,i} <r_a$ and $\lambda_{g,1}<\lambda_{g,2}$. 

The variable $\lambda_{g,i}$ is used to signify a minimum distance between UAV-$i$ and the GS \cite[eq. (3)]{chetlur2017downlink}. Furthermore, using the PDF of $f_{d_{g,i}}(w)$, the reliability of downlink NOMA for UAV communications for bivariate Rician shadowed fading channels can now be analyzed through stochastic geometry approaches.

\subsection{Received Signal at the Downlink UAVs}

At the downlink UAVs, NOMA necessitates the adoption of the II and SIC detectors. As UAV-1 is closer to the GS ($\lambda_{g,1}<\lambda_{g,2}$), we assume an imperfect SIC detector at UAV-1 and an II detector at UAV-2. Let $x_{gs}=\sqrt{a_{gs,1}}x_{gs,1}+\sqrt{a_{gs,2}}x_{gs,2}$ be the transmitted signal from the GS, where $x_{gs,i}$ is the signal-of-interest (SOI) for UAV-$i$ and $a_{gs,i}$ is the power allocation factor at UAV-$i$ satisfying $a_{gs,1}+a_{gs,2}=1$. Then, the received signal at downlink UAV-$i$ can be written as $y_{i} = \sqrt{\frac{P_t}{d_{g,i}^{2}}}h_{g,i}x_{gs} + w_{i}$, where $h_{g,i}$ denotes the channel between the GS and UAV-$i$ and $w_{i}$ denotes the additive white Gaussian noise (AWGN) at UAV-$i$ with zero-mean and variance $\eta_i$.


\section{Bivariate Rician Shadowed Distribution} \label{sec_bi_var}
In this section, the Bivariate Rician shadowed fading model is introduced. We begin by noting that the Bivariate Rician shadowed distributed random variable (RV) $H_{k}$, $k \in \{1,2\}$ is modeled as \cite{lopez2018bivariate}:
\begin{eqnarray} \label{rv_bi_rician_shad}
H_k = \sigma\sqrt{1-\rho}X_k + \sigma\sqrt{\rho}X_0 + Z,
\end{eqnarray}
where $X_k, k \in \{0,1,2\}$ are Gaussian RVs with zero mean and variance $\frac{1}{2}$ and $0 \leq \rho \leq 1$ is the cross correlation coefficient. Denoting $E\{\bullet\}$ as the statistical expectation operator, we note that $E\big\{ \big(\sigma\sqrt{1-\rho}X_k + \sigma\sqrt{\rho}X_0\big)^2 \big\} = \sigma^2$. Finally, $Z$ is a Nakagami-m distributed RV with shaping parameter $m\geq 0.5$ and $E\{|Z|^2\}=\Omega_N$.

The RV $|H_{k}|$ follows a bivariate Rician shadowed distribution with $E\{|H_k|^2\}=\sigma^2(1+K)$ and Rician factor $K=\frac{\Omega_N}{\sigma^2}$. In \cite[eq. (4)]{lopez2018bivariate}, the joint PDF $f_{R_1,R_2}(r_1,r_2)$ of $R_k = |H_k|$ is presented as: 
\begin{eqnarray} \label{pdf_bi_rician_shad}
\hspace{-0.3cm} f_{R_1,R_2}(r_1,r_2) & = & \frac{8 (\frac{m \rho}{m \rho + K})^m}{\sigma^6 \rho (1-\rho)^2} r_1 r_2 \exp\bigg(-\frac{r_1^2 + r_2^2}{\sigma^2(1-\rho)}\bigg) \nonumber \\
 & & \hspace{-2.7cm} \times \int_0^\infty x \exp\bigg( \frac{-(1+\rho)}{\sigma^2 \rho(1-\rho)}x^2 \bigg) I_0\bigg( \frac{2r_1x}{\sigma^2(1-\rho)} \bigg) I_0\bigg( \frac{2r_2x}{\sigma^2(1-\rho)} \bigg) \nonumber \\
& & \hspace{1cm} \times {}_1{F_1}\bigg(m,1;\frac{K}{\sigma^2 \rho(\rho m + K)}x^2\bigg) dx_,
\end{eqnarray}
where $I_{0}\left(\bullet\right)$ is the modified Bessel function of the first kind with zero order \cite[eq. (9.6.10)]{abramowitz1964handbook} and ${}_1{F_1}(\bullet)$ is the confluent Hypergeometric function \cite{gradshteyn2014table}.

In the current form, evaluating commonly used metrics, e.g., outage probability, using the joint PDF expression in (\ref{pdf_bi_rician_shad}) requires the use of numerical methods. Instead, we now present a closed-form expression for $f_{R_1,R_2}(r_1,r_2)$ in the following Lemma:
\begin{lemma} \label{lemma_pwr_srs_pdf_bi_rician_shad}
The closed-form expression for $f_{R_1,R_2}(r_1,r_2)$ can be expressed as the following power series:
\begin{eqnarray} \label{pwr_srs_pdf_bi_rician_shad}
& & \hspace{-1cm} f_{R_1,R_2}(r_1,r_2) \nonumber \\
& \hspace{-0.5cm} \approx & \hspace{-0.25cm} \sum_{k=0}^{K_{tr,1}} \sum_{i=0}^{k} \sum_{n=0}^{i} \alpha(k,i,n) r_1^{2n+1} r_2^{2(i-n)+1} \exp\bigg( -\frac{r_1^2 + r_2^2}{\sigma^2 (1-\rho)} \bigg)_,
\end{eqnarray}
\end{lemma}
where $\alpha(k,i,n) = \frac{8(m)_{k-i}\big( \frac{K}{\sigma^2 \rho (\rho m + K)} \big)^{k-i} \big(\frac{m\rho}{m\rho+K}\big)^m}{\Gamma^2(n+1)\Gamma^2(i-n+1)[\sigma^2 (1-\rho)]^{2i} (1)_{k-i} (k-i)! \sigma^6 \rho (1-\rho)^2} \\ \times \frac{k!}{2\big(\frac{1+\rho}{\sigma^2 \rho (1-\rho)}\big)^{k+1}}$, $K_{tr,j}$ for $j \in \{1,2\}$ is the truncation order, and $(a)_k=\frac{\Gamma(a+k)}{\Gamma(a)}$ is the Pochhammer symbol \cite[eq. (6.1.22)]{abramowitz1964handbook}.
\begin{IEEEproof}
The proof is provided in Appendix \ref{pdf_lemma_proof}.
\end{IEEEproof}

From (\ref{pwr_srs_pdf_bi_rician_shad}), one can also obtain the closed-form marginal CDF $F_{R_i}(\gamma_i)$ in the following Lemma:
\begin{lemma} \label{lemma_pwr_srs_cdf_bi_rician_shad}
The closed-form expressions for $F_{R_1}(\gamma_1)$ and $F_{R_2}(\gamma_2)$ can be expressed as:
\begin{eqnarray} 
F_{R_1}(\gamma_1) & \approx & \sum_{k=0}^{K_{tr,1}} \sum_{i=0}^{k} \sum_{n=0}^{i} \sum_{j=0}^{K_{tr,2}} \alpha(k,i,n) G(j,n,i-n,\gamma_1)_, \label{pwr_srs_cdf_r1} \\
F_{R_2}(\gamma_2) & \approx & \sum_{k=0}^{K_{tr,1}} \sum_{i=0}^{k} \sum_{n=0}^{i} \sum_{j=0}^{K_{tr,2}} \alpha(k,i,n) G(j,i-n,n,\gamma_2)_, \label{pwr_srs_cdf_r2}
\end{eqnarray}
\end{lemma}
where $G(j,l,q,\gamma) = \frac{(-1)^j \gamma^{2(l+j+1)} q!}{j![\sigma^2(1-\rho)]^{j-q-1}4(l+j+1)}$.
\begin{IEEEproof}
The proof is provided in Appendix \ref{cdf_lemma_proof}.
\end{IEEEproof}

Using the closed-form expressions in Lemma \ref{lemma_pwr_srs_cdf_bi_rician_shad} as the basis, one obtains the closed-form outage probability expressions of the downlink UAVs with NOMA, which is shown in the next section.

\section{Outage Probability at the Downlink UAVs} \label{sec_outage}

The outage probability expressions for NOMA at the downlink UAVs are presented in this section. The outage probability expressions for OMA at the downlink UAVs are also presented as a benchmark. Let the transmission rate of the GS be defined as $R^{i}_{gs}$ for $i \in \{NOMA, OMA\}$, where we let $R_{gs}^{NOMA}=\frac{1}{2}R_{gs}^{OMA}$ for a fair comparison between NOMA and OMA.

\subsection{NOMA Outage Probability}

With UAV-1 in close proximity to the GS as compared to UAV-2 ($\lambda_{g,1}<\lambda_{g,2}$), an imperfect SIC detector and an II detector is considered at UAV-1 and UAV-2, respectively. At UAV-$i$, let $R_i=|h_{g,i}|$ and $\gamma_i^{NOMA*}$ be the instantaneous received signal envelope of the SOI and the normalized NOMA threshold, respectively. Specifically, $\gamma_1^{NOMA*} = \sqrt{\frac{\gamma_1^{NOMA}}{P_{g,1}[a_{gs,1}-(1-a_{gs,1})\beta\gamma_1^{NOMA}]}}$ and $\gamma_2^{NOMA*} = \sqrt{\frac{\gamma_2^{NOMA}}{P_{g,2}[a_{gs,2}-(1-a_{gs,2})\gamma_2^{NOMA}]}}$, where $\gamma_i^{NOMA}=2^{R_{gs}^{NOMA}}-1$ is the NOMA threshold, $P_{g,i}=P_t/\eta_i$ is the normalized transmit power, and $0 \leq \beta \leq 1$ denotes the strength of the residual interference due to imperfect SIC.

Using the above definitions, the outage event for NOMA at UAV-$i$ is defined as $\mathcal{O}_{i}^{NOMA} = \Big\{ R_{i}, d_{g,i} : R_i < \gamma_i^{NOMA*}d_{g,i}\Big\}$. Next, the closed-form outage probability expression for UAV-$i$ is presented in the following theorem.

\begin{theorem} \label{theorem_P_out_down_uav}
The NOMA outage probability at downlink UAV-$i$ for $i \in \{1,2\}$ is
\begin{eqnarray} \label{P_out_down_uav}
Pr\big(\mathcal{O}_{1}^{NOMA}) & \approx & \sum_{k=0}^{K_{tr,1}} \sum_{i=0}^{k} \sum_{n=0}^{i} \sum_{j=0}^{K_{tr,2}} \alpha(k,i,n) \nonumber \\
 & & \hspace{-0.2cm} \times G(j,n,i-n,\gamma_1^{NOMA*}) \overline{G}(\lambda_{g,1},n+j)_, \\ 
Pr\big(\mathcal{O}_{2}^{NOMA}) & \approx & \sum_{k=0}^{K_{tr,1}} \sum_{i=0}^{k} \sum_{n=0}^{i} \sum_{j=0}^{K_{tr,2}} \alpha(k,i,n) \nonumber \\
 & & \hspace{-0.7cm} \times G(j,i-n,n,\gamma_2^{NOMA*}) \overline{G}(\lambda_{g,2},i-n+j)_,
\end{eqnarray}
\end{theorem}
where $\overline{G}(\lambda,k) = \frac{(D_{alt}^2 + \lambda^2 + r_a^2)^{k+2} - (D_{alt}^2 + \lambda^2)^{k+2}}{r_a^2 (k+2)}$.
\begin{IEEEproof}
Theorem \ref{theorem_P_out_down_uav} can be obtained by first evaluating the conditional outage probability $Pr\big(\mathcal{O}_{i}^{NOMA}|d_{g,i})$ using Lemma \ref{lemma_pwr_srs_cdf_bi_rician_shad} before averaging over the PDF $f_{d_{g,i}}(w)$.
\end{IEEEproof}

Using Theorem \ref{theorem_P_out_down_uav}, an evaluation of downlink NOMA in multi-UAV networks under a stochastic geometry framework is now possible.

\subsection{OMA Outage Probability}

As discussed earlier, univariate fading models are suitable for point-to-point UAV links in OMA. In this spirit, we employ the univariate Rician shadowed fading model in \cite{tan2018ricianShad} for UAV channel modeling. Let $X_i = P_{g,i}|h_{g,i}|^2$ be the instantaneous received signal power of the SOI, where $X_i$ is a Rician shadowed distributed RV with Rician $K$ factor $K_{X{_i}}$ and shaping parameter $m_{X_{i}}$. Also, let $\gamma_i^{OMA}=2^{R_{gs}^{OMA}}-1$ be the OMA threshold. Then, the outage event for OMA at UAV-$i$ is defined as $\mathcal{O}_{i}^{OMA} = \Big\{ h_{g,i}, d_{g,i} : X_i < \gamma_i^{OMA}d_{g,i}^2\Big\}$. Using \cite[eq. (10)]{tan2018ricianShad}, along with the same approach in Theorem \ref{theorem_P_out_down_uav}, the OMA outage probability at UAV-$i$ can be obtained as:
\begin{eqnarray} \label{P_out_down_oma_uav}
Pr\big(\mathcal{O}_{i}^{OMA}) & \hspace{-0.25cm} \approx & \hspace{-0.25cm} \sum_{k=0}^{K_{tr,1}} \sum_{i=0}^{k} \overline{\alpha}(k,i,P_{g,i},\gamma_i^{OMA}) \overline{G}(\lambda_{g,1},n+j)_,
\end{eqnarray}
where $\overline{\alpha}(k,i,P_{g,i},\gamma_i^{OMA}) = (-1)^{k-i} \Big(\frac{m_{X_{i}}}{K_{X{_i}}+m_{X_{i}}}\Big)^{m_{X_{i}}} \frac{(m_{X_{i}})_i}{\Gamma^2(i+1)} \Big(\frac{K_{X{_i}}}{K_{X{_i}}+m_{X_{i}}}\Big)^{i} \Big(\frac{1+K_{X{_i}}}{P_{g,i}}\Big)^{k+1} \frac{\gamma^{k+1}}{(k-i)!(k+1)}$.

\section{Numerical Results} \label{sec_num_res}

\begin{table}[]
\centering
\caption{Simulation Parameters}
\label{table:sim_param}
\begin{tabular}{ll}
\hline
\textbf{Parameter(s)} 																					& \textbf{Value(s)}																		\\  \hline \hline
Truncation orders $\{K_{tr,1}, K_{tr,2}\}$											& $\{30,10\}$																					\\
Rician $K$ factors																							& 10\text{dB} \cite[Table V]{matolak2017air_suburban} for $\sigma=1$	\\
Cross correlation coefficient $\rho$														& 0.5																									\\
Transmission rate $R_{gs}^{OMA}$																& 0.1 b/s/Hz 																					\\ 
Power allocation factor $a_{gs,1}$															& 0.5 			 																					\\ 
Residual imperfect SIC interference $\beta$											& 0.01																								\\
Radius $r_a$																										&	4 km																								\\ 
Altitude $D_{alt}$																							&	0.2 km																							\\
Minimum distance $\{\lambda_{g,1}, \lambda_{g,2}\}$							&	$\{2\text{ km}, 3\text{ km}\}$ 											\\	\hline
\end{tabular}
\end{table}

In this section, the validation of the joint PDF expression and the outage probabilities at the downlink UAVs are presented. We also present Monte Carlo simulations conducted with $10^{7}$ samples using the simulation parameters in Table \ref{table:sim_param}.

\begin{figure} [] 
\centering
\includegraphics [width=0.925\columnwidth]{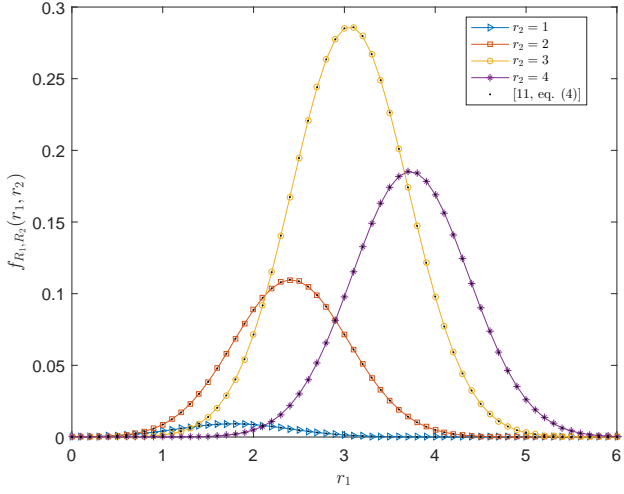} 
\caption{PDF comparison between the expression in \cite[eq. (4)]{lopez2018bivariate} and the closed-form expression in (\ref{pwr_srs_pdf_bi_rician_shad}) for $m=10$ and $K_{tr,1}=150$.}
\label{fig:pdf_comp}
\end{figure}

In Fig. \ref{fig:pdf_comp}, the closed-form expression for the PDF $f_{R_1,R_2}(r_1,r_2)$ in (\ref{pwr_srs_pdf_bi_rician_shad}) is compared against the expression in (\ref{pdf_bi_rician_shad}), which is obtained from \cite[eq. (4)]{lopez2018bivariate}. Evidently, (\ref{pwr_srs_pdf_bi_rician_shad}) is shown to be in very close agreement with \cite[eq. (4)]{lopez2018bivariate}. Furthermore, as $m \to \infty$, the closed-form expression in (\ref{pwr_srs_pdf_bi_rician_shad}) can be used to model a bivariate Rician fading PDF. 

\begin{figure} [] 
\centering
\includegraphics [width=0.925\columnwidth]{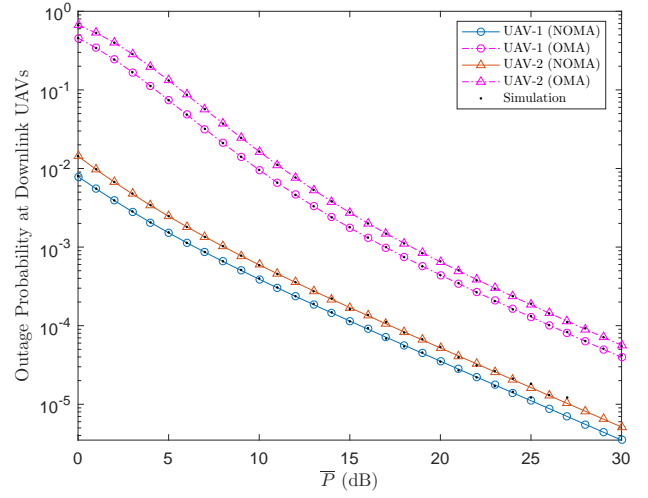} 
\caption{Outage probability comparison of NOMA and OMA at the downlink UAVs for $P_{g,1}=P_{g,2}=\overline{P}$.}
\label{fig:downlink_outage}
\end{figure}
The outage probabilities of the downlink UAVs are plotted in Fig. \ref{fig:downlink_outage}. It is observed that the downlink UAVs with NOMA exhibit lower outage probability than OMA. The reason for such a trend is because the transmission rate of NOMA is set at half of OMA, i.e., $R_{gs}^{NOMA}=\frac{1}{2}R_{gs}^{OMA}$, for a fair comparison. More interestingly, for NOMA, the outage probabilities of the downlink UAVs are not interference-limited at high $P_t$ regimes, despite the presence of residual interference ($\beta$). It is also observed that, for both NOMA and OMA, UAV-1 attains lower outage probability than UAV-2. Such a trend implies that the imperfect SIC at UAV-1 is able to attain a lower outage probability than the II detector at UAV-2 as the former is closer to the GS than the latter ($\lambda_{g,1}<\lambda_{g,2}$).

\begin{figure} []
\centering
\includegraphics [width=0.925\columnwidth]{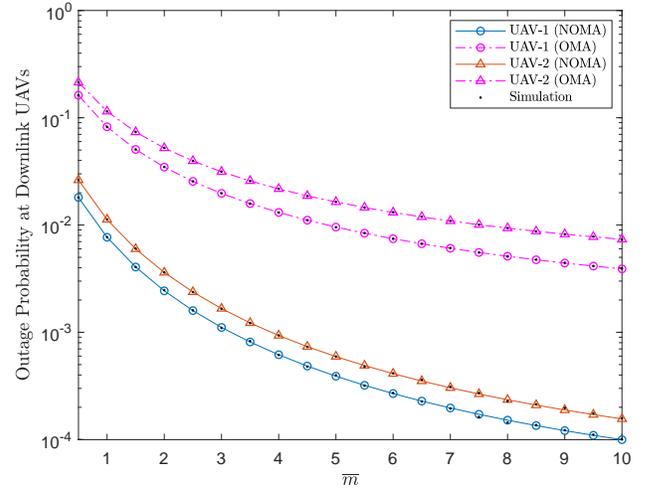} 
\caption{Impact of shadowing on NOMA and OMA at the downlink UAVs for $P_{g,1}=P_{g,2}=10$dB and $m=m_{X_1}=m_{X_2}=\overline{m}$.}
\label{fig:shad_downlink_outage}
\end{figure}
The impact of shadowing on the downlink UAVs is shown in Fig. \ref{fig:shad_downlink_outage}. It is seen that NOMA attains lower outage probability than OMA, even when shadowing is severe, e.g., $\overline{m}\leq 1$. Furthermore, it is also observed that the outage probability of NOMA decays faster than OMA as shadowing reduces ($\overline{m} \to \infty$). As the trends in Fig. \ref{fig:shad_downlink_outage} are plotted with Rician $K$ factors of 10dB, further analytical analysis will be needed to determine the combined impact of the Rician $K$ factor and shadowing parameters on NOMA outage probability.

\section{Conclusion} \label{sec_conclusion}

In this paper, downlink non-orthogonal multiple access (NOMA) is analyzed for multi-UAV networks as a potential solution to improve spectrum utilization. Utilizing a power series approach, closed-form expressions are derived for the joint PDF and marginal CDF in the context of a bivariate Rician shadowed fading model. With the marginal CDF expressions, closed-form outage probability expressions are also derived for a multi-UAV network with downlink NOMA under a stochastic geometry framework. An analysis of the outage probability for the multi-UAV network showed downlink NOMA attaining lower outage probability than OMA. Also, NOMA is shown to be less affected by the impact of shadowing than OMA.

\appendices
\section{Proof of Lemma \ref{lemma_pwr_srs_pdf_bi_rician_shad}} \label{pdf_lemma_proof}
We first begin by noting that $I_0\big( \frac{2r_ix}{\sigma^2(1-\rho)} \big)$ for $i=1,2$ in (\ref{pdf_bi_rician_shad}) can be represented as the following power series \cite[eq. (9.6.10)]{abramowitz1964handbook}:
\begin{eqnarray} \label{mod_bessel_pwr_series}
I_0\bigg( \frac{2r_ix}{\sigma^2(1-\rho)} \bigg) = \sum_{n=0}^{\infty} \frac{(1/4)^n}{n! \Gamma(n+1)} \bigg( \frac{2r_ix}{\sigma^2 (1-\rho)} \bigg)^{2n} \hspace{-0.2cm} = \sum_{n=0}^{\infty} C_i(n). \hspace{-0.2cm}
\end{eqnarray}

Then, using the Cauchy product theorem \cite[eq. (0.316)]{gradshteyn2014table}, $\prod_{i=1}^2 I_0\big( \frac{2r_ix}{\sigma^2(1-\rho)} \big)$ in (\ref{pdf_bi_rician_shad}) becomes:
\begin{eqnarray} \label{prod_mod_bessel_pwr_series}
\prod_{i=1}^2 I_0\bigg( \frac{2r_ix}{\sigma^2(1-\rho)} \bigg) \approx \sum_{k=0}^{\infty} \sum_{n=0}^{k} C_1(n) C_2(k-n) \approx \sum_{k=0}^{K_{tr,1}} A(k),
\end{eqnarray}
where $A(k) = \sum_{n=0}^{k} \frac{(1/4)^k (2r_1)^{2n}(2r_2)^{2(k-n)}}{\Gamma^2(n+1)\Gamma^2(k-n+1)[\sigma^2 (1-\rho)]^{2k}}x^{2k}$.

Next, ${}_1{F_1}\big(m,1;\frac{K}{\sigma^2 \rho(\rho m + K)}x^2\big)$ in (\ref{pdf_bi_rician_shad}) is also expressed as the following power series \cite[eq. (13.1.2)]{abramowitz1964handbook}:
\begin{eqnarray} \label{conf_hyp_pwr_series}
{}_1{F_1}\bigg(m,1;\frac{K}{\sigma^2 \rho(\rho m + K)}x^2\bigg) \approx \sum_{i=0}^{\infty} B(i),
\end{eqnarray}
where $B(i) = \frac{(m)_i}{i!(1)_i}\big( \frac{K}{\sigma^2 \rho (m\rho+K)} \big)^i x^{2i}$. Using (\ref{prod_mod_bessel_pwr_series}) and (\ref{conf_hyp_pwr_series}), along with the Cauchy product theorem \cite[eq. (0.316)]{gradshteyn2014table}, $\prod_{i=1}^2 I_0\big( \frac{2r_ix}{\sigma^2(1-\rho)} \big) {}_1{F_1}\big(m,1;\frac{K}{\sigma^2 \rho(\rho m + K)}x^2\big)$ in (\ref{pdf_bi_rician_shad}) can be expressed as:
\begin{eqnarray} \label{prod_mod_bessel_conf_hyp_pwr_series}
 & & \hspace{-1cm} \prod_{i=1}^2 I_0\bigg( \frac{2r_ix}{\sigma^2(1-\rho)} \bigg) {}_1{F_1}\bigg(m,1;\frac{K}{\sigma^2 \rho(\rho m + K)}x^2\bigg) \nonumber\\
 & \approx & \sum_{k=0}^{K_{tr,1}} \sum_{i=0}^{k} A(i) B(k-i) \nonumber \\
 & \approx & \sum_{k=0}^{K_{tr,1}} \sum_{i=0}^{k} \sum_{n=0}^{i} \frac{(\frac{1}{4})^i (2r_1)^{2n}(2r_2)^{2(i-n)}}{\Gamma^2(n+1)\Gamma^2(i-n+1)[\sigma^2(1-\rho)]^{2i}(1)_{k-i}} \nonumber \\
 & & \hspace{2.3cm} \times \frac{(m_{k-i})}{(k-i)!} \bigg(\frac{K}{\sigma^2 \rho (\rho m + K)}\bigg)^{k-i}x^{2k}.
\end{eqnarray}

Substituting (\ref{prod_mod_bessel_conf_hyp_pwr_series}) into (\ref{pdf_bi_rician_shad}) and utilizing the fact that $\int_0^{\infty} x^{2k+1} \exp\Big( \frac{-(1-\rho)}{\sigma^2 \rho(1-\rho)}x^2 \Big)dx = \frac{k!}{2}\Big(\frac{1-\rho}{\sigma^2 \rho(1-\rho)}\Big)^{-(k+1)}$ \cite[eq. (3.461.3)]{gradshteyn2014table}, one obtains the expression in (\ref{pwr_srs_pdf_bi_rician_shad}). This completes the proof.

\section{Proof of Lemma \ref{lemma_pwr_srs_cdf_bi_rician_shad}} \label{cdf_lemma_proof}
Starting with $R_1$, we note that $\exp\big(\frac{-x^2}{A}\big) \approx \sum_{j=0}^{K_{tr,2}} \frac{(-1)^j}{j!A^j}x^{2j}$ \cite[eq. (1.211.3)]{gradshteyn2014table}. Then, the marginal CDF $F_{R_1}(\gamma_1)$ can be obtained from (\ref{pwr_srs_pdf_bi_rician_shad}) as follows:
\begin{eqnarray}
F_{R_1}(\gamma_1) & \approx & \int_{0}^{\infty} \int_{0}^{\gamma_1} \sum_{k=0}^{K_{tr,1}} \sum_{i=0}^{k} \sum_{n=0}^{i} \alpha(k,i,n) r_1^{2n+1} r_2^{2(i-n)+1} \nonumber \\
 & & \hspace{2.5cm} \times \exp\bigg( -\frac{r_1^2 + r_2^2}{\sigma^2 (1-\rho)} \bigg) dr_1 dr_2 \nonumber \\
 & \hspace{-2.6cm} \approx & \hspace{-1.5cm} \sum_{k=0}^{K_{tr,1}} \sum_{i=0}^{k} \sum_{n=0}^{i} \sum_{j=0}^{K_{tr,2}} \alpha(k,i,n) \frac{(-1)^j (\gamma_1)^{2(n+j+1)}}{j![\sigma^2(1-\rho)]^j 2(n+j+1)} \nonumber \\
 & & \hspace{0.5cm} \times \int_{0}^{\infty} r_2^{2(i-n)+1} \exp\bigg( -\frac{r_2^2}{\sigma^2 (1-\rho)} \bigg) dr_2. \label{proof_pwr_srs_cdf_bi_rician_shad}
\end{eqnarray}

Applying $\int_0^{\infty} x^{2k+1} \exp\Big( \frac{-(1-\rho)}{\sigma^2 \rho(1-\rho)}x^2 \Big)dx = \frac{k!}{2}\Big(\frac{1-\rho}{\sigma^2 \rho(1-\rho)}\Big)^{-(k+1)}$ \cite[eq. (3.461.3)]{gradshteyn2014table} to evaluate the integral in (\ref{proof_pwr_srs_cdf_bi_rician_shad}) yields (\ref{pwr_srs_cdf_r1}). For the marginal CDF $F_{R_2}(\gamma_2)$, applying the same approach yields (\ref{pwr_srs_cdf_r2}). This completes the proof.

\ifCLASSOPTIONcaptionsoff
  \newpage
\fi

\bibliographystyle{IEEEtran}
\bibliography{IEEEabrv,ref}

\end{document}